\documentclass[showpacs,aps,twocolumn]{revtex4}
\usepackage{epsfig}
\usepackage{graphicx}
\usepackage{amsmath,amssymb,amsfonts}
\usepackage{amsmath}
\usepackage{array}
\usepackage{url}
\usepackage{multirow}
\usepackage{float}

\usepackage{booktabs}  
\usepackage{longtable} 

\usepackage{lineno}
\usepackage{xspace}
\usepackage{xcolor}
\usepackage[normalem]{ulem} 

\newcommand{\bef}{\begin{figure}}
\newcommand{\eef}{\end{figure}}
\newcommand{\bc}{\begin{center}}
\newcommand{\ec}{\end{center}}

\newcommand{\be}{\begin{equation}}
\newcommand{\ee}{\end{equation}}
\newcommand{\bea}{\begin{eqnarray}}
\newcommand{\eea}{\end{eqnarray}}

\def\ba{\begin{eqnarray}}
\def\ea{\end{eqnarray}}

\definecolor{darkblue}{RGB}{0,0,196}
\usepackage[colorlinks=true,linkcolor=darkblue,citecolor=darkblue,urlcolor=darkblue]{hyperref}

\begin{document}
\title{Anisotropic hadronic matter in external magnetic fields}

\author{Dushmanta Sahu}
\email{Dushmanta.Sahu@cern.ch}
\affiliation{Instituto de Ciencias Nucleares, UNAM, Apartado Postal 70-543, Coyoacán, 04510, México City, México}
\author{Jayanta Dey}
\email{jayanta@theor.jinr.ru}
\affiliation{Bogoliubov Laboratory of Theoretical Physics, Joint Institute for Nuclear Research,
6 Joliot-Curie St., Dubna, 141980, Russia}

\begin{abstract}
Strong magnetic fields generated in non-central heavy-ion collisions break rotational symmetry and can imprint anisotropy on the produced quantum chromodynamics (QCD) matter. Within a hadron resonance gas framework in the presence of an external magnetic field, we investigate the resulting anisotropy in the equation of state, isothermal compressibility, speed of sound, and conserved-charge fluctuations. We find that the electric-charge susceptibility ratio $\chi_4^Q/\chi_2^Q$ develops a pronounced directional dependence, driven primarily by the Landau-level structure of charged pions. For physically relevant magnetic fields, this anisotropy reaches $\sim10$--$15\%$ at $T\simeq0.16$ GeV. The ratio of electric-charge susceptibility can be expressed as $\chi_4^Q/\chi_2^Q=C_4^Q/C_2^Q=\kappa_Q\sigma_Q^2$, where $C_2^Q$ and $C_4^Q$ are the second- and fourth-order cumulants, and $\kappa_Q$ and $\sigma_Q$ are the kurtosis and variance of the event-by-event net-electric-charge distribution. The reaction-plane dependence of this ratio provides an experimentally accessible signature of magnetic-field-induced anisotropy. Our results identify higher-order electric-charge fluctuations as a promising probe of the imprint of strong electromagnetic fields on the matter produced in heavy-ion collisions at RHIC and the LHC.
\end{abstract}
\maketitle

\section{Introduction}
\label{sec:intro}

The properties of strongly interacting matter in the presence of external magnetic fields have attracted considerable attention due to their relevance for relativistic heavy-ion collisions, neutron stars, and magnetars~\cite{Kharzeev:2007jp,Skokov:2009qp,McLerran:2013hla}. In non-central heavy-ion collisions at the Relativistic Heavy Ion Collider (RHIC) and the Large Hadron Collider (LHC), the fast-moving spectator protons generate extremely strong, albeit short-lived, magnetic fields with typical strengths reaching $eB \sim 0.02$--$0.4~\mathrm{GeV}^2$~\cite{Skokov:2009qp,McLerran:2013hla,Tuchin:2013ie}. Although these magnetic fields decay rapidly in the absence of medium effects, the conducting quark-gluon plasma can induce electric currents through Faraday induction as the magnetic field evolves, thereby slowing its decay and extending its lifetime. Consequently, the magnetic field may persist sufficiently long to influence the subsequent evolution of the medium and modify its thermodynamic and transport properties. Thus, understanding the response of QCD matter to external magnetic fields has become an important part of the broader effort to characterize strongly interacting matter under extreme conditions~\cite{Endrodi:2024cqn,Ding:2015ona,Fukushima:2011jc,Pradhan:2021vtp,Sahoo:2023vkw,Sahu:2025tmb}.

A fundamental consequence of an external magnetic field is the explicit breaking of rotational symmetry. In isotropic matter, the pressure is a scalar quantity, whereas in a magnetized medium it naturally separates into longitudinal ($P_{\parallel}$) and transverse ($P_{\perp}$) components with respect to the direction of the magnetic field~\cite{Karmakar:2019tdp,Chaudhuri:2022oru,Ferraris:2025fva,Goswami:2023eol}. This pressure anisotropy modifies the equation of state and the mechanical response of the medium, leading to directional dependence in quantities such as the speed of sound ($c_s^{2}$) and isothermal compressibility ($\kappa_{T}$)~\cite{Ferraris:2025fva,Goswami:2023eol}. More generally, external magnetic fields can induce anisotropies in thermodynamic and response properties, including conserved-charge fluctuations and transport coefficients~\cite{Singh:2024emy, Singh:2023ues, Dey:2020awu, Dey:2019vkn, Dey:2019axu}. Since these quantities enter the hydrodynamic evolution of the medium, quantifying their magnetic-field dependence and anisotropy is essential for realistic descriptions of magnetized QCD matter.

The influence of magnetic fields on QCD thermodynamics has been extensively investigated within effective descriptions of deconfined matter, particularly the Nambu--Jona-Lasinio (NJL) and Polyakov-loop extended NJL (PNJL) models~\cite{Chaudhuri:2022oru,Goswami:2023eol,Andersen:2014xxa,Endrodi:2019whh,Avancini:2019wed}. These studies have explored the modification of the equation of state, speed of sound, isothermal compressibility, magnetic susceptibility, and the interplay between magnetic fields and chiral symmetry restoration. Related investigations have also been carried out within perturbative one-loop QCD\cite{Rath:2017fdv,Karmakar:2019tdp} and in the context of magnetized neutron stars~\cite{Dexheimer:2016yqu}. On the other hand, the Hadron Resonance Gas (HRG) model has been widely employed to investigate the thermodynamics of hadronic matter in magnetic fields, including the equation of state, magnetization, and the effects of hadronic interactions through excluded-volume and van der Waals extensions~\cite{Endrodi:2013cs,Kadam:2019rzo,Sahoo:2023vkw}. Despite these developments, comparatively little attention has been devoted to the anisotropy of conserved-charge fluctuations in magnetized hadronic matter. To the best of our knowledge, a systematic HRG framework that connects anisotropic thermodynamics with higher-order conserved-charge cumulants has not yet been established.

A central open question in this context is not merely whether the magnetic field imprints anisotropy on the medium, but whether that imprint survives to an experimentally accessible, field-induced signature. Momentum-space anisotropies such as the elliptic flow ($v_2$) are dominated by the initial collision geometry and the subsequent hydrodynamic response; an observed anisotropy in $v_2$ cannot be cleanly attributed to the magnetic field without disentangling these much larger conventional sources. Conserved-charge fluctuations offer a conceptually distinct handle: event-by-event cumulants of net-proton, net-charge, and net-kaon multiplicity distributions, serving as experimental proxies for net-baryon, electric-charge, and net-strangeness fluctuations, respectively, are routinely measured in searches for the QCD critical point and studies of the QCD phase transition~\cite{STAR:2025zdq,STAR:2022vlo, Luo:2017faz,Sarkar:2014fja}. A reaction-plane-dependent measurement of these cumulants; comparing fluctuations in-plane and out-of-plane relative to the (predominantly perpendicular) magnetic field, probes the field's imprint on the medium's response functions directly. Establishing what magnitude of directional dependence an ideal hadronic medium alone would predict for such an observable is therefore a necessary baseline for interpreting any future measurement.

The HRG model provides a good description of QCD thermodynamics below the crossover temperature, where hadronic degrees of freedom dominate~\cite{Karsch:2003vd,Ratti:2010kj}. It provides a natural framework for investigating magnetic-field effects in the hadronic phase. In an external magnetic field, the transverse motion of charged hadrons is quantized into Landau levels, whereas neutral hadrons do not experience this orbital coupling to the field~\cite{Endrodi:2013cs}. This modification of the phase space of charged hadrons can consequently induce directional dependence in thermodynamic observables, providing a well-defined hadronic baseline for more microscopic descriptions of magnetized QCD matter.

In this work, we use a self-consistent thermodynamic framework for an ideal HRG in the presence of an external magnetic field. Charged hadrons are treated through Landau quantization, while vacuum contributions are incorporated using a renormalized thermodynamic potential. Starting from the resulting anisotropic equation of state, we derive the longitudinal and transverse pressures, magnetization, isothermal compressibility, and speed of sound. We further extend the formalism to conserved-charge fluctuations by constructing direction-dependent generalized susceptibilities. This paper is organized as follows. In Sec.~\ref{formalism}, we present the theoretical framework of the magnetized ideal HRG model and derive the anisotropic thermodynamic quantities together with the formalism for anisotropic fluctuation observables. Numerical results and their physical implications are discussed in Sec.~\ref{results}. Finally, we summarize our conclusions in Sec.~\ref{summary}.

\section{Formalism}
\label{formalism}

We begin with the partition function and the modifications induced by the magnetic field via Landau quantization in an ideal HRG model. Subsequently, we derive the fundamental thermodynamic quantities and extend the formalism to describe the anisotropic nature of the system, including direction-dependent pressures, compressibility, speed of sound, and fluctuations.

\subsection{Ideal HRG in an External Magnetic Field}

We consider an ideal gas of hadrons and resonances subjected to a uniform, static external magnetic field, \(\vec{B} = B\hat{z}\), aligned along the \(z\)-axis. The grand canonical partition function for this non-interacting system is given by the sum over all hadron species \(i\),
\begin{equation}
    \ln Z = \sum_i \ln Z_i,
\end{equation}
where the partition function for the \(i\)-th particle species is~\cite{Karsch:2003zq},
\begin{equation}
    \ln Z_i = \pm Vg_i \int \frac{d^3p}{(2\pi)^3} \ln \left\{ 1 \pm \exp\left[-(E_i - \mu_i)/T\right]\right\}.
\end{equation}
Here, \(V\) is the volume, \(T\) is the temperature, \(g_i\) is the degeneracy factor, and \(E_i\) is the single-particle energy. The upper (lower) sign corresponds to fermions (bosons). The chemical potential \(\mu_i\) is defined as \(\mu_i = B_i\mu_B + S_i\mu_S + Q_i\mu_Q\), where \(B_i\), \(S_i\), and \(Q_i\) are the baryon number, strangeness, and electric charge of the particle, and $\mu_{B}$, $\mu_{S}$ and $\mu_{Q}$ are the baryon chemical potential, strange chemical potential and electric charge chemical potential, respectively.

The presence of the magnetic field quantizes the transverse momentum of charged particles into Landau levels. The dispersion relation for a charged hadron (\(Q_i \neq 0\)) becomes~\cite{Endrodi:2013cs},
\begin{equation}\label{eq:dispersion}
    E_{c,i}(p_z, k, s_z) = \sqrt{p_z^2 + m_i^2 + 2|Q_i|B\left(k + \frac{1}{2} - s_z\right)},
\end{equation}
where \(p_z\) is the momentum along the field direction, \(k = 0, 1, 2, \ldots\) is the Landau level index, \(s_z\) is the spin projection. For neutral hadrons (\(Q_i = 0\)), the energy spectrum remains unaffected,
\begin{equation}
    E_{n,i}(p) = \sqrt{p^2 + m_i^2}.
\end{equation}

Consequently, the phase space integral for charged particles transforms as follows~\cite{Endrodi:2013cs,Pradhan:2021vtp},
\begin{equation}
    \int \frac{d^3p}{(2\pi)^3} \rightarrow \frac{|Q_i|B}{2\pi^2} \sum_{k=0}^{\infty} \sum_{s_z} \int_0^\infty dp_z.
    \label{eq5}
\end{equation}

Using the partition function, the primary thermodynamic quantities for each particle species are derived as follows;

\begin{equation}
    P_i(T,\mu_i,B) = \pm \frac{|Q_i|B T}{2\pi^2} \sum_{k=0}^{\infty}\sum_{s_z}
    \int_0^\infty \!\! dp_z\,
    \ln\!\left\{1 \pm e^{-(E_i-\mu_i)/T}\right\}.
    \label{eq6}
\end{equation}

\begin{equation}
    \epsilon_i(T,\mu_i,B) = \frac{|Q_i|B}{2\pi^2} \sum_{k=0}^{\infty}\sum_{s_z}
    \int_0^\infty \!\! dp_z\,
    \frac{E_i}{e^{(E_i-\mu_i)/T}\pm 1}.
    \label{eq7}
\end{equation}

\begin{equation}
    n_i(T,\mu_i,B) = \frac{|Q_i|B}{2\pi^2} \sum_{k=0}^{\infty}\sum_{s_z}
    \int_0^\infty \!\! dp_z\,
    \frac{1}{e^{(E_i-\mu_i)/T}\pm 1}.
    \label{eq8}
\end{equation}

\begin{align}
    s_i(T,\mu_i,B) =\;& \pm \frac{|Q_i|B}{2\pi^2} \sum_{k=0}^{\infty}\sum_{s_z}
    \int_0^\infty \!\! dp_z\,
    \Bigg[ \ln\!\left\{1\pm e^{-(E_i-\mu_i)/T}\right\} \nonumber \\
    &\pm\, \frac{(E_i-\mu_i)/T}{e^{(E_i-\mu_i)/T}\pm1} \Bigg].
    \label{eq9}
\end{align}

Here, $P$, $\epsilon$, $n$ and $s$ are the pressure, energy density, number density, and entropy density, respectively. The total value for each quantity is obtained by summing over all hadron species. Note that hadrons are composite particles; particularly for resonances, in the case of contribution to the magnetization, it becomes important to consider the correct gyromagnetic ratio of each hadron. In other words, the contribution from the anomalous magnetic moment (AMM) due to the correction in the Lande $g$ factor of the composite hadron becomes important at large magnetic fields~\cite{Samanta:2025mrq}. However, in the current work, we consider magnetic fields on the order of a few \(m_{\pi }^{2}\); we have calculated that the effect of the AMM is very small and thus neglect it in the results.

\subsection{Anisotropic Thermodynamics in a Magnetic Field}
\label{sec:anisotropic_thermo}

The presence of an external magnetic field breaks rotational symmetry and
leads to different pressures parallel and perpendicular to the field
direction~\cite{Karmakar:2019tdp}. In the grand canonical ensemble, the longitudinal pressure is
related to the partition function through,
\begin{equation}
    P_\parallel(T,\{\mu_i\},B)
    = \frac{T}{V}\ln Z(T,\{\mu_i\},B)
    = \frac{T}{V}\sum_i \ln Z_i(T,\mu_i,B).
\end{equation}

The transverse pressure receives an additional contribution from the
magnetization of the medium and is given by,
\begin{equation}
    P_\perp
    = P_\parallel-BM,
\end{equation}
where the magnetization can be expressed directly in terms of the partition
function as,
\begin{equation}
    M
    = \frac{T}{V}
    \left(\frac{\partial \ln Z}{\partial B}\right)_
    {T,\{\mu_i\}}
    = \sum_i M_i.
\end{equation}

Thus, the anisotropy between the transverse and longitudinal pressures is
determined by the magnetic-field dependence of the HRG partition function,
which originates from the Landau quantization of charged hadrons. It should be noted that the pressure anisotropy in a polarizable medium under an external magnetic field depends on the treatment of the energy-momentum tensor (EMT), varying depending on whether the system is open or strictly conserves the EMT. For instance, Ref.~\cite{Bali:2013esa} demonstrated that the pressure in a specific direction is determined by the response of the thermodynamic potential to spatial compression along that axis. When evaluating the derivative under a constant magnetic field, the pressure remains isotropic. Conversely, evaluating it under a constant magnetic flux creates pressure anisotropy, which is exactly what we consider here.

\subsection{Vacuum Renormalization}
\label{sec:vacuum}

In the presence of an external magnetic field, the thermodynamic potential receives both thermal and vacuum contributions. Accordingly, the total pressure can be written as
\begin{equation}
P_{\mathrm{total}}(T,eB)
=
P_{\mathrm{th}}(T,eB)
+
\Delta P_{\mathrm{vac}}(T=0,eB).
\end{equation}
The vacuum contribution is ultraviolet divergent and must therefore be renormalized. Following the magnetic-field-independent regularization and renormalization procedure discussed in Ref.~\cite{Sahoo:2023vkw}, the field-dependent vacuum contribution can be written in terms
of derivatives of the Hurwitz zeta function. The ultraviolet divergences are removed by subtracting the $B=0$ contribution and absorbing the remaining divergence into the renormalization of the electromagnetic field and electric charge.

For spin-\(\tfrac{1}{2}\) particles, the resulting renormalized vacuum pressure is,
\begin{equation}
\begin{split}
\Delta P_{\mathrm{vac}}^{r}(S=\tfrac{1}{2},B)
=
\frac{(|Q|B)^2}{4\pi^2}
\Bigg[
&\zeta'(-1,x)
+\frac{x}{2}\ln x
-\frac{x^2}{2}\ln x \\[2pt]
&+\frac{x^2}{4}
-\frac{1}{12}\ln x
-\frac{1}{12}
\Bigg],
\end{split}
\end{equation}
where
\begin{equation}
x=\frac{m^2}{2|Q|B},
\end{equation}
and \(\zeta'(-1,x)\) denotes the derivative of the Hurwitz zeta function with respect to its first argument.

For spin-\(0\) particles,
\begin{equation}
\begin{split}
\Delta P_{\mathrm{vac}}^{r}(S=0,B)
=
-\frac{(|Q|B)^2}{8\pi^2}
\Bigg[
&\zeta'(-1,x+\tfrac{1}{2})
-\frac{x^2}{2}\ln x \\[2pt]
&+\frac{x^2}{4}
+\frac{\ln x+1}{24}
\Bigg].
\end{split}
\end{equation}

For spin-\(1\) particles,
\begin{equation}
\begin{aligned}
\Delta P_{\mathrm{vac}}^{r}(S=1,B)
=
&-\frac{3(|Q|B)^2}{8\pi^2}
\Bigg[
\zeta'(-1,x-\tfrac{1}{2}) \\[2pt]
&+\frac{x+\tfrac{1}{2}}{3}\ln(x+\tfrac{1}{2}) \\[2pt]
&+\frac{2}{3}(x-\tfrac{1}{2})\ln(x-\tfrac{1}{2}) \\[2pt]
&-\frac{x^2}{2}\ln x
+\frac{x^2}{4}
-\frac{7}{24}(\ln x+1)
\Bigg].
\end{aligned}
\end{equation}

The total renormalized vacuum contribution is obtained by summing the corresponding field-dependent contributions over all charged hadrons included in the hadron resonance gas,
\begin{equation}
\Delta P_{\mathrm{vac}}^{r}(B)
=
\sum_{i\in\mathrm{charged}}
\Delta P_{\mathrm{vac},i}^{r}(B),
\end{equation}
with the appropriate expression determined by the spin of each hadron.

\subsection{Anisotropic Compressibility and Speed of Sound}
\label{sec:aniso_observables}

The isothermal compressibility, \(\kappa_T = -\frac{1}{V}\left(\frac{\partial V}{\partial P}\right)_T\), measures the relative change in volume with pressure at constant temperature. Using \(V = N/n\), it can be rewritten in terms of number density \(n\). We present these relations generically for an arbitrary conserved-charge chemical potential \(\mu_X\), where \(X\) denotes either the net-proton (baryon-like) or electric-charge sector. For the associated number density \(n_X\), the longitudinal and transverse components are:
\begin{align}
{\kappa_{T}}_{\parallel}
&=
\frac{1}{n_X}
\left(\frac{\partial n_X}{\partial P_{\parallel}}\right)_{T,B}
=
\frac{1}{n_X^2}
\left(\frac{\partial n_X}{\partial \mu_X}\right)_{T,B}, \\[4pt]
{\kappa_{T}}_{\perp}
&=
\frac{1}{n_X}
\left(\frac{\partial n_X}{\partial P_{\perp}}\right)_{T,B}
=
\frac{1}{n_X}
\frac{
\left(\frac{\partial n_X}{\partial \mu_X}\right)_{T,B}
}{
n_X -
B\left(\frac{\partial M}{\partial \mu_X}\right)_{T,B}
}.
\end{align}
The additional term in \({\kappa_T}_\perp\) arises from the dependence of the transverse pressure on magnetization.

The adiabatic speed of sound squared is defined as \(c_s^2 = \left(\frac{\partial P}{\partial \epsilon}\right)_{s/n}\). In the presence of a magnetic field and chemical potential, this derivative must be evaluated keeping the entropy per particle (\(s/n\)) and the magnetic field \(B\) constant, leading to a non-trivial expression \cite{Sahoo:2023vkw}:
\begin{equation}
c_{s,a}^{2}(T,\mu,B)
=
\frac{
\left(\frac{\partial P^a}{\partial T}\right)_{\mu,B}
+
\left(\frac{\partial P^a}{\partial \mu}\right)_{T,B}
\left(\frac{\partial \mu}{\partial T}\right)_{s/n,B}
}{
\left(\frac{\partial \epsilon}{\partial T}\right)_{\mu,B}
+
\left(\frac{\partial \epsilon}{\partial \mu}\right)_{T,B}
\left(\frac{\partial \mu}{\partial T}\right)_{s/n,B}
},
\end{equation}
where $a$ represents longitudinal ($\parallel$) and transverse ($\perp$) components.
The derivative \(\left(\frac{\partial \mu}{\partial T}\right)_{s/n,B}\) under the constant \(s/n\) and fixed \(B\) constraint is given by:
\begin{equation}
\left(\frac{\partial \mu}{\partial T}\right)_{s/n,B}
=
\frac{
s\left(\frac{\partial n}{\partial T}\right)_{\mu,B}
-
n\left(\frac{\partial s}{\partial T}\right)_{\mu,B}
}{
n\left(\frac{\partial s}{\partial \mu}\right)_{T,B}
-
s\left(\frac{\partial n}{\partial \mu}\right)_{T,B}
}.
\end{equation}
These equations must be solved numerically. The anisotropy enters through the different pressure components, \(P_\parallel\) and \(P_\perp\), used in the numerator.

\subsection{Anisotropic Fluctuation Observables}
\label{sec:fluctuations}

Since conserved-charge fluctuations are generated from derivatives of the pressure with respect to the corresponding chemical potential, the pressure anisotropy naturally induces anisotropic cumulants and generalized susceptibilities.

By denoting the longitudinal pressure as $P~ (= P_\parallel$), the generalized susceptibilities in the longitudinal direction are defined as
\begin{equation}
{\chi_n}_{\parallel}
=
\frac{\partial^n}{\partial(\mu_X/T)^n}
\left(\frac{P}{T^4}\right)_{T,B}.
\end{equation}

Using the transverse pressure,
\begin{equation}
P_\perp = P - B\mathcal{M},
\end{equation}
the transverse susceptibilities become
\begin{equation}
{\chi_n}_\perp
=
\frac{\partial^n}{\partial(\mu_X/T)^n}
\left(
\frac{P - B\mathcal{M}}{T^4}
\right)_{T,B}.
\end{equation}

Since the external magnetic field is independent of the chemical potential,
\begin{equation}
{\chi_n}_\perp
=
{\chi_n}_\parallel
-
\frac{B}{T^4}
\frac{\partial^n\mathcal{M}}
{\partial(\mu_X/T)^n}.
\end{equation}

It is convenient to define the mixed magneto-susceptibilities
\begin{equation}
\chi_n^{B}
=
\frac{\partial^n\mathcal{M}}
{\partial(\mu_X/T)^n}
=
\frac{\partial^{n+1}P}
{\partial B\,\partial(\mu_X/T)^n},
\end{equation}
which quantify the response of conserved-charge fluctuations to the external magnetic field.

The magnetic-field-induced anisotropy of fluctuations can be quantified through the normalized observable,
\begin{equation}
A_n
=
\frac{{\chi_n}_\parallel - {\chi_n}_\perp}
{{\chi_n}_\parallel} = \frac{\Delta \chi_n}{{\chi_n}_\parallel},
\end{equation}
which vanishes in the absence of a magnetic field and directly measures the anisotropy generated by magnetization. For our study, we essentially plot $|A_n|$. Here, we calculate electric charge susceptibility ($\chi^Q$) for vanishing chemical potential $\mu_Q = 0$.

\begin{figure*}
    \centering
\includegraphics[width=0.95\linewidth]{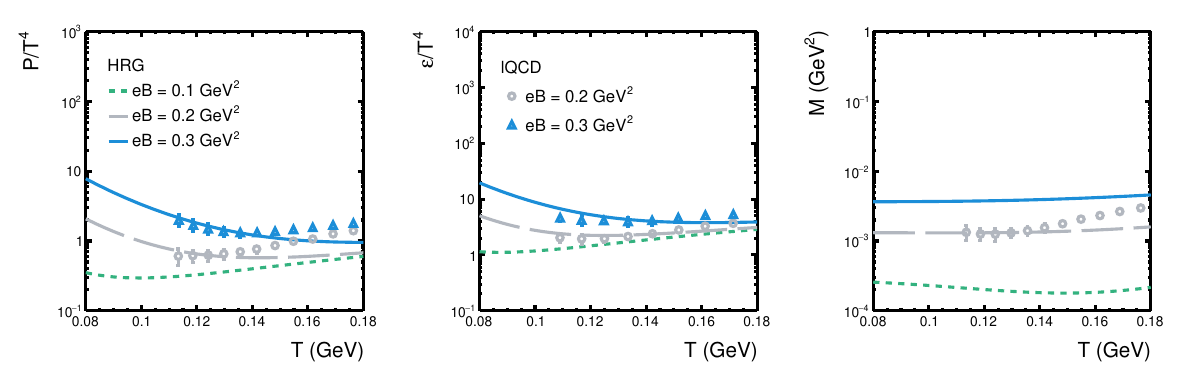}
    \caption{Left panel shows the normalized total pressure $P/T^4$ as a function of $T$. Middle panel shows the normalized energy density $\varepsilon/T^4$ as a function of $T$, with the same line and marker conventions. Right panel shows the total magnetization $M$ as a function of $T$. The solid lines represent the HRG results for different magnetic field strengths and the markers denote lattice QCD data from Ref.~\cite{Bali:2014kia}.}
    \label{fig1}
\end{figure*}

\begin{figure*}
    \centering
\includegraphics[width=0.95\linewidth]{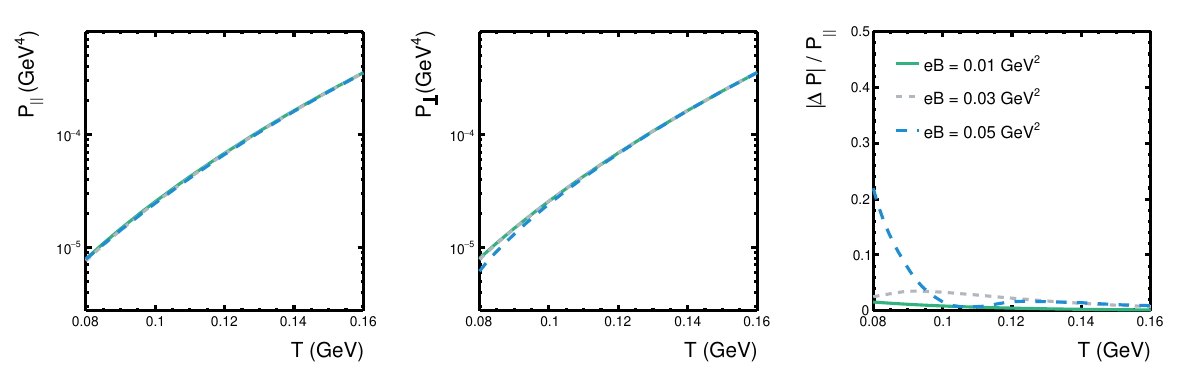}
\caption{Left panel shows the longitudinal pressure $P_{\parallel}$ as a function of $T$. Middle panel shows the transverse pressure $P_{\perp}$ as a function of $T$, with the same line conventions. Right panel shows the relative pressure anisotropy $|\Delta P|/P_{\parallel}$ as a function of $T$.}
    \label{fig2}
\end{figure*}

\begin{figure*}
    \centering
\includegraphics[width=0.95\linewidth]{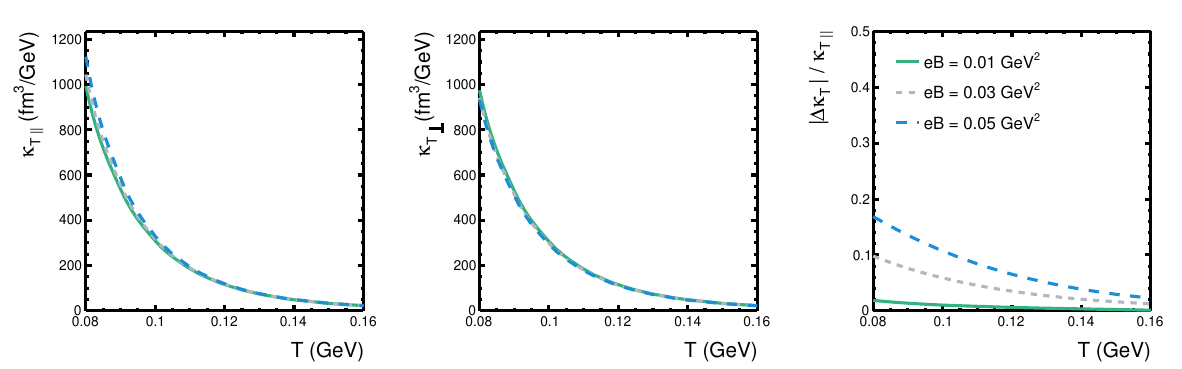}
\caption{Left panel shows the longitudinal isothermal compressibility $\kappa_{T_{\parallel}}$ as a function of $T$. Middle panel shows the transverse isothermal compressibility $\kappa_{T_{\perp}}$ as a function of $T$, with the same line conventions. Right panel shows the relative compressibility anisotropy $|\Delta\kappa_T|/\kappa_{T_{\parallel}}$ as a function of $T$.}
    \label{fig3}
\end{figure*}

\begin{figure*}
    \centering
\includegraphics[width=0.95\linewidth]{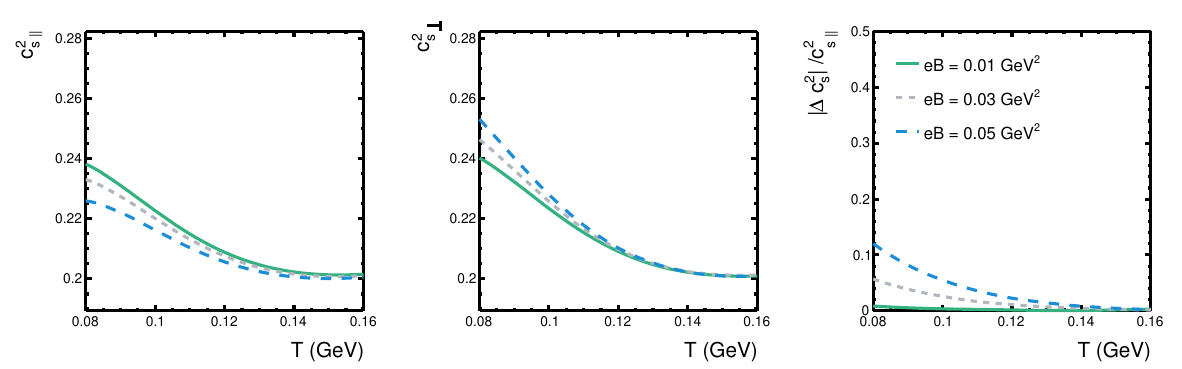}
\caption{Left panel shows the longitudinal squared speed of sound $c_{s_{\parallel}}^{2}$ as a function of $T$. Middle panel shows the transverse squared speed of sound $c_{s_{\perp}}^{2}$ as a function of $T$, with the same line conventions. Right panel shows the relative speed-of-sound anisotropy $|\Delta c_s^2|/c_{s_{\parallel}}^{2}$ as a function of $T$.}
    \label{fig4}
\end{figure*}

\section{Results and Discussion}
\label{results}

Table~\ref{tab:particles} lists the hadron species included in the present calculation, together with their mass, electric charge, spin and baryon number. For a consistency check, we have plotted the scaled longitudinal ($\parallel$) pressure $P/T^{4}$, scaled energy density $\varepsilon/T^{4}$ and the magnetization $M$ as a function of temperature for three different $eB$ values in the left, middle and right panel of Fig.\ref{fig1}, respectively. These particular $eB$ values are taken so as to have a direct comparison with the lattice QCD results.

\begin{table}[h]
\centering
\caption{Hadron species included in the HRG calculation. $Q$, $S$, and $B$ denote electric charge, spin, and baryon number, respectively.}
\label{tab:particles}
\begin{tabular}{|l|c|c|c|c|}
\hline
\hline
Particle & Mass (GeV) & $Q$ & $S$ & $B$ \\
\hline
\multicolumn{5}{|l|}{\textit{Pseudoscalar mesons}} \\
\hline
$\pi^0$   & 0.135 & 0  & 0 & 0 \\
$\pi^+$   & 0.140 & +1 & 0 & 0 \\
$\pi^-$   & 0.140 & $-1$ & 0 & 0 \\
$K^0$     & 0.498 & 0  & 0 & 0 \\
$K^+$     & 0.494 & +1 & 0 & 0 \\
$K^-$     & 0.494 & $-1$ & 0 & 0 \\
$\eta$    & 0.548 & 0  & 0 & 0 \\
$\eta'$   & 0.958 & 0  & 0 & 0 \\
\hline
\multicolumn{5}{|l|}{\textit{Vector mesons}} \\
\hline
$\rho^0$   & 0.775 & 0  & 1 & 0 \\
$\rho^+$   & 0.776 & +1 & 1 & 0 \\
$\rho^-$   & 0.776 & $-1$ & 1 & 0 \\
$\omega$   & 0.782 & 0  & 1 & 0 \\
$K^{*+}$   & 0.892 & +1 & 1 & 0 \\
$K^{*-}$   & 0.892 & $-1$ & 1 & 0 \\
$K^{*0}$   & 0.896 & 0 & 1 & 0 \\
\hline
\multicolumn{5}{|l|}{\textit{Baryons}} \\
\hline
$p$        & 0.938 & +1 & 1/2 & +1 \\
$n$        & 0.940 & 0  & 1/2 & +1 \\
$\bar{p}$  & 0.938 & $-1$ & 1/2 & $-1$ \\
$\bar{n}$  & 0.940 & 0  & 1/2 & $-1$ \\
$\Sigma^+$ & 1.189 & +1 & 1/2 & +1 \\
$\Sigma^0$ & 1.192 & 0  & 1/2 & +1 \\
$\Sigma^-$ & 1.197 & $-1$ & 1/2 & +1 \\
$\Lambda$  & 1.116 & 0  & 1/2 & +1 \\
$\Xi^-$    & 1.321 & $-1$ & 1/2 & +1 \\
$\Xi^0$    & 1.315 & 0  & 1/2 & +1 \\
\hline
\hline
\end{tabular}
\end{table}

The left panel of Fig.~\ref{fig2} shows the parallel pressure component as a function of temperature for three different values of external magnetic fields, $eB$ = 0.01, 0.03 and 0.05 $\rm GeV^{2}$. Magnetic fields generated in high-energy collisions decay rapidly; consequently, the field strength is substantially suppressed by the hadronic stage. To investigate whether such a weak magnetic field can induce measurable anisotropies in final-state observables, we adopt these correspondingly small field values in our estimations. The middle panel shows the perpendicular pressure and the right panel shows the pressure anisotropy in a hadron gas for the given external magnetic field. Here, the parallel component have magnetic field dependency only through Landau quantization, whereas, transverse component is affected from medium's magnetization too. Both pressure components, $P_\parallel$ and $P_\perp$, increase monotonically with $T$. This behavior is driven by the increase in particle occupation numbers, as a broader region of thermodynamic phase space becomes thermally accessible with increasing temperature. Though the $eB$ dependence on the individual components are negligible, from the right panel we can see that the pressure anisotropy increases with increasing magnetic-field strength, reaching up to $20\%$ at lower temperatures for $eB=0.05~\rm GeV^2$. Note that the anisotropy for any observable ($\mathcal{O}$) is defined by $|\Delta \mathcal{O}|/\mathcal{O}_\parallel$, with $\Delta \mathcal{O} = \mathcal{O}_\parallel - \mathcal{O}_\perp$. For all considered magnetic-field strengths, however, the relative pressure anisotropy decreases with increasing temperature and becomes negligible at higher $T$. As the temperature increases, the thermal contribution to the pressure grows rapidly, thereby reducing the relative importance of the magnetic contribution $MB$. Consequently, the pressure anisotropy decreases toward zero at high temperatures. \\

For the isothermal compressibility, shown in Fig.~\ref{fig3}, we observe a systematic pattern where the parallel component $\kappa_T^\parallel$ consistently exceeds the perpendicular component $\kappa_T^\perp$ across all temperatures and magnetic field strengths. Also, both components of compressibility decrease with increase in temperature, suggesting that at higher $T$ regime, the system gradually moves towards a more ideal scenario, approaching incompressibility or stiffer equation of state. The right panel of Fig.~\ref{fig3} shows the anisotropy in isothermal compressibility, a higher $eB$ case showing a higher anisotropy. Here also, the anisotropy decreases with increase in temperature for all magnetic field values.

For the squared speed of sound, shown in Fig.~\ref{fig4}, we observe a systematic difference between the parallel and perpendicular components. The parallel component ${c_s{^2}}_{\parallel}$ remains consistently lower than the perpendicular component ${c_s^{2}}_{\perp}$ across the considered temperature and magnetic-field ranges, unlike the isothermal compressibilty. This indicates that the external magnetic field induces a directional dependence in the propagation of pressure perturbations through the medium. The right panel of Fig.~\ref{fig4} shows the anisotropy in the squared speed of sound, which increases with increasing magnetic-field strength. Similar to the behavior observed for the pressure and isothermal compressibility, the anisotropy decreases with increasing temperature for all considered magnetic-field values, indicating that thermal effects progressively suppress the relative impact of the magnetic field on the equation of state.

The anisotropic equation of state obtained in the present work has direct implications for relativistic magnetohydrodynamic descriptions of heavy-ion collisions. In a magnetized medium, the energy--momentum tensor is no longer characterized by a single isotropic pressure, but instead contains distinct longitudinal and transverse pressures, relative to the magnetic-field direction. Consequently, the pressure gradients responsible for driving the collective expansion of the medium become direction dependent, leading to anisotropic acceleration and modified propagation of hydrodynamic perturbations. Since the speed of sound and isothermal compressibility are likewise anisotropic, the thermodynamic response of the hadronic medium to local density fluctuations is expected to differ along and perpendicular to the magnetic field. These effects provide important microscopic inputs for realistic magnetohydrodynamic simulations, where the anisotropic hadronic equation of state can influence the late-time evolution of the medium after hadronization. Although the rapidly decaying magnetic field may reduce the magnitude of these effects at late times, the present calculations establish a quantitative hadronic baseline that can be incorporated into future magnetohydrodynamic studies of heavy-ion collisions.

\begin{figure*}
    \centering
\includegraphics[width=0.95\linewidth]{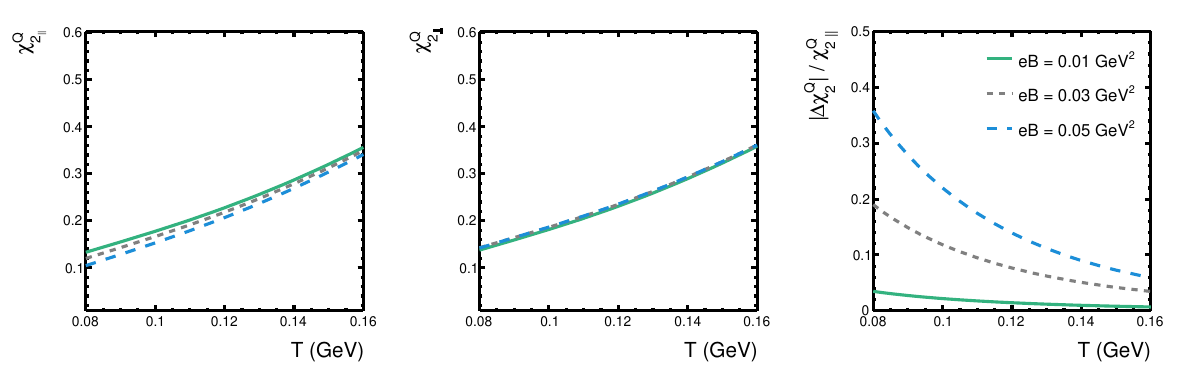}
\caption{Left panel shows the longitudinal second-order electric-charge susceptibility ${\chi^{Q}_{2}}_{\parallel}$ as a function of $T$. Middle panel shows the transverse second-order electric-charge susceptibility ${\chi_{2}^{Q}}_{\perp}$ as a function of $T$, with the same line conventions. Right panel shows the relative susceptibility anisotropy as a function of $T$.}
    \label{fig5}
\end{figure*}

\begin{figure*}
    \centering
\includegraphics[width=0.95\linewidth]{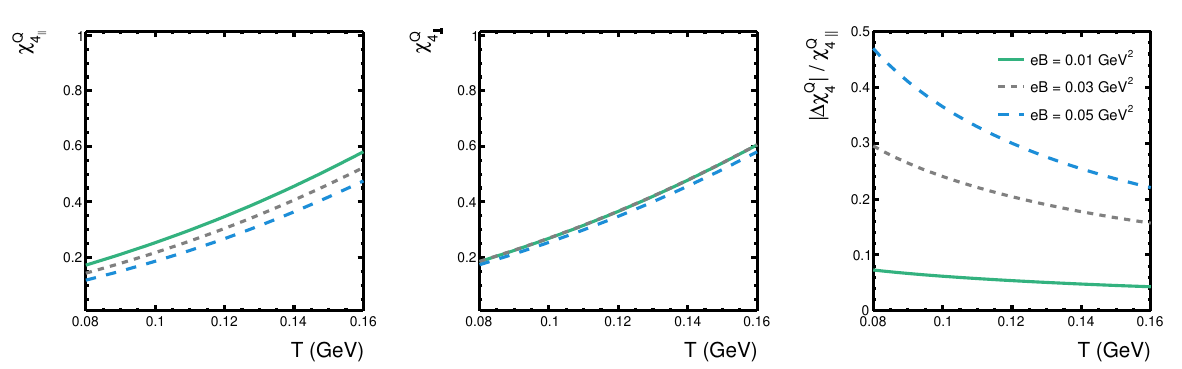}
\caption{Left panel shows the longitudinal fourth-order electric-charge susceptibility ${\chi_{4}^{Q}}_{\parallel}$ as a function of $T$. Middle panel shows the transverse fourth-order electric-charge susceptibility ${\chi_{4}^{Q}}_{\perp}$ as a function of $T$, with the same line conventions. Right panel shows the relative susceptibility anisotropy as a function of $T$.}
    \label{fig6}
\end{figure*}

\begin{figure*}
    \centering
\includegraphics[width=0.95\linewidth]{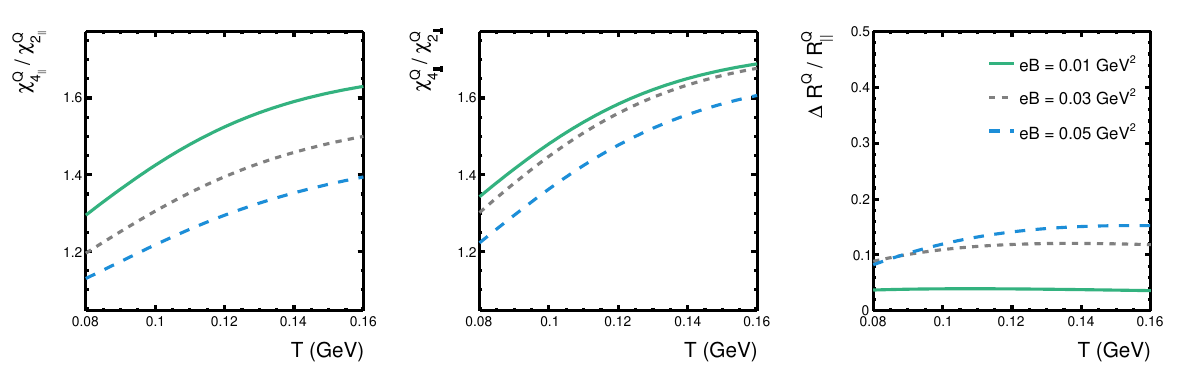}
\caption{Left panel shows the longitudinal electric-charge cumulant ratio $R^{Q}_{\parallel} = {\chi_4^{Q}}_{\parallel}/{\chi_2^{Q}}_{\parallel}$ as a function of $T$. Middle panel shows the transverse ratio $R^{Q}_{\perp} = {\chi_4^{Q}}_{\perp}/{\chi_2^{Q}}_{\perp}$ as a function of $T$, with the same line conventions. Right panel shows the relative ratio anisotropy as a function of $T$.}
    \label{fig7}
\end{figure*}

In Fig.~\ref{fig5}, we show the second-order electric-charge susceptibility for both parallel, perpendicular direction, and its anisotropy. Both components of $\chi_2^Q$ increase with temperature, reflecting the increasing thermal excitation of electrically charged hadrons and the corresponding enhancement of electric-charge fluctuations. The right panel of Fig.~\ref{fig5} shows the anisotropy in the second-order electric-charge susceptibility, which increases systematically with increasing $eB$. Similar to the behavior observed for the pressure, isothermal compressibility, and speed of sound, the relative anisotropy decreases with increasing temperature for all considered magnetic-field strengths, indicating that the thermal contribution progressively reduces the relative impact of the magnetic field on electric-charge fluctuations. One thing to note here is that the electric-charge susceptibility receives contributions from the full spectrum of charged hadrons and is dominated, at the temperatures considered here, by the lightest charged species, namely pions. Unlike protons, pions are bosons, and their Landau-level threshold energy, is shifted upward by the magnetic field even at the lowest level ($k=0$), in contrast to the field-independent fermionic threshold. Since $eB\sim0.01$--$0.05~\rm GeV^2$ is comparable to or exceeds $m_\pi^2$ ($\approx0.02~\rm GeV^2$), this shift is not a small perturbation and produces a significant magnetic modification of the pion phase space. Importantly, this modification affects different susceptibility orders differently. 

For the fourth-order electric-charge susceptibility, shown in Fig.~\ref{fig6}, we observe that both components of $\chi_4^Q$ increase with temperature, reflecting the increasing contribution of thermally excited electrically charged hadrons to higher-order electric-charge fluctuations. The right panel of Fig.~\ref{fig6} shows the anisotropy in the fourth-order electric-charge susceptibility, which increases systematically with increasing $eB$. Similar to the behavior observed for the second-order susceptibility, pressure, isothermal compressibility, and speed of sound, the relative anisotropy decreases with increasing temperature for all considered magnetic-field strengths, indicating that the thermal contribution progressively reduces the relative impact of the magnetic field on higher-order electric-charge fluctuations.

Finally, for the ratio of the fourth- to second-order electric-charge susceptibilities, shown in Fig.~\ref{fig7}, we observe a clear directional dependence between the parallel and perpendicular components. The anisotropy in $\chi_4^Q/\chi_2^Q$ becomes more pronounced with increasing magnetic-field strength, indicating that the magnetic field affects the second- and fourth-order electric-charge fluctuations differently. In contrast to the individual susceptibilities, the ratio provides a particularly sensitive measure of the magnetic-field-induced anisotropy, as the magnetic corrections to $\chi_2^Q$ and $\chi_4^Q$ do not cancel completely. The right panel of Fig.~\ref{fig7} shows that the anisotropy in $\chi_4^Q/\chi_2^Q$ increases systematically with increasing $eB$, reaching values of $\sim10$--$15\%$ near $T\simeq0.16~\rm GeV$ for the considered magnetic-field strengths. Unlike other quantities studied here, the relative anisotropy doesn't decrease with increasing temperature, indicating that the directional sensitivity of the cumulant ratio to the external magnetic field remains substantial even as thermal effects become dominant.

This observation suggests that anisotropic cumulant ratios can provide a sensitive probe of magnetic-field-induced effects. In particular, $\chi_4^Q/\chi_2^Q$ provides a natural target for reaction-plane-dependent fluctuation measurements, as the ideal HRG predicts a sizable directional dependence arising from the Landau-level structure of charged pions. Experimentally, the second- and higher-order net-electric-charge cumulants can be measured separately for particles emitted in-plane and out-of-plane with respect to the reaction plane, allowing the construction of
\begin{equation}
A_n^Q=
\frac{{\chi_n^{Q}}_{\rm out}-{\chi_n^{Q}}_{\rm in}}
{{\chi_n^{Q}}_{\rm out}}.\nonumber
\end{equation}

Since the average magnetic field is perpendicular to the reaction plane, the
out-of-plane (in-plane) direction corresponds to the $\parallel$
($\perp$) direction used in our calculation. $A_n^Q$ is therefore related to the relative anisotropy plotted here. A nonzero $A_n^Q$ would provide a constraint on magnetic-field-induced anisotropy in the hadronic phase. Such measurements, particularly of $\chi_4^Q/\chi_2^Q$, would establish an important magnetic-field baseline for interpreting conserved-charge fluctuations and for searches for the QCD critical point and the nature of the phase transition.

\section{Summary}
\label{summary}

We study the anisotropy produced in an ideal hadron resonance gas in an external magnetic field, incorporating Landau quantization of charged hadrons and a renormalized vacuum contribution. We find that the magnetic field induces a pronounced directional dependence in the hadronic equation of state, reflected in the pressure, isothermal compressibility, and speed of sound. The resulting anisotropy increases systematically with the field strength and is most significant at lower temperatures, while thermal effects progressively reduce its relative magnitude at higher temperatures. These results establish that an external magnetic field can leave a sizable imprint on the bulk thermodynamics of the hadronic phase. We further demonstrate that this anisotropy extends to conserved-charge fluctuations, with the electric-charge susceptibilities exhibiting a clear directional dependence. Most notably, the susceptibility ratio $\chi_4^Q/\chi_2^Q$ develops a genuine magnetic-field-induced anisotropy of $\sim10\text{--}15\%$ near $T \simeq 0.16\text{ GeV}$ across the considered field strengths. Because conserved charge fluctuations map directly to experimental event-by-event cumulants via the volume-independent relation $\chi_4^Q/\chi_2^Q = \kappa_Q \sigma_Q^2$, the persistence of this anisotropic signature is of prime phenomenological importance. It provides an experimentally measurable observable that is sensitive to the directional structure of the magnetized medium.

We note, however, that an experimentally observed anisotropy in net-electric-charge cumulants would not be uniquely attributable to the magnetic field, as local charge conservation and chiral magnetic effects can also generate reaction-plane-dependent charge fluctuations. Our results therefore establish a quantitative hadronic baseline for magnetic-field-induced anisotropies in both bulk thermodynamics and conserved-charge fluctuations. In particular, reaction-plane-dependent measurements of net-electric-charge cumulants, and especially $\kappa_Q \sigma_Q^2$, offer a concrete experimental avenue for constraining the imprint of strong electromagnetic fields on the hadronic medium produced in non-central heavy-ion collisions. Such measurements can provide an essential input for future magnetohydrodynamic descriptions and for disentangling magnetic-field-induced backgrounds from other sources of anisotropy in fluctuation observables.

\section*{Acknowledgement}
DS acknowledges the support from the postdoctoral fellowship of the DGAPA UNAM.
JD acknowledges support from the JINR Postdoctoral Programme.\\

\end{document}